\documentclass[A4]{aa}
\usepackage{graphicx}
\begin{document}
\title{\emph{XMM--Newton} Observations of the Ultraluminous Nuclear\\X-ray Source in
M33\thanks{Based on public observations obtained with \emph{XMM-Newton}, an ESA
science mission with instruments and contributions directly funded by ESA Member
States and the USA (NASA).}}

\author{L. Foschini\inst{1}, J. Rodriguez\inst{2,3}, Y. Fuchs\inst{2}, L. C. Ho\inst{4},
M. Dadina\inst{1},\\G. Di Cocco\inst{1}, T. J.-L. Courvoisier\inst{3,5}, G. Malaguti\inst{1}}

\offprints{L. Foschini \email{foschini@bo.iasf.cnr.it}}

\institute{Istituto di Astrofisica Spaziale e Fisica Cosmica (IASF) del
CNR, Sezione di Bologna, Via Gobetti 101, 40129 Bologna (Italy)
\and
CEA Saclay, DSM/DAPNIA/SAp (CNRS FRE 2591) F--91191 Gif--sur--Yvette Cedex, France
\and
INTEGRAL Science Data Centre, Chemin d'\'Ecogia 16, CH--1290 Versoix, Switzerland
\and
The Observatories of the Carnegie Institution of Washington, 813 Santa Barbara Street, Pasadena, CA 91101 (USA)
\and
Observatory of Geneva, 51 chemins des Mailettes, 1290 Sauverny, Switzerland
}

\date{Received 10 September 2003; accepted 3 December 2003}

\abstract{We present observations with \emph{XMM-Newton} of M33 X-8, the ultraluminous
X-ray source ($L_{0.5-10\,\rm keV}\approx 2\times 10^{39}$ erg/s) closest to the centre of the galaxy.
The best-fit model is similar to the typical model of Galactic black holes in very high state.
Comparison with previous observations indicates that the source is still in a
very high state after about 20 years of observations.  No state transition has
been observed even during the present set of \emph{XMM-Newton} observations. 
We estimate the lower limit of the mass of the black hole $>6 \,M_{\odot}$, but 
with proper parameters taking into account different effects, the best estimate becomes 
$12\,M_{\odot}$. Our analysis favours the hypothesis that M33 X--8 is a stellar mass
black hole candidate, in agreement with the findings of other authors. In addition, 
we propose a different model where the high luminosity of the source
is likely to be due to orientation effects of the accretion disc and anisotropies in the
Comptonized emission.
}

\authorrunning{L. Foschini et al.}
%\titlerunning{XMM-Newton observation}
\maketitle

\section{Introduction}
About 20 years ago, with the early observations of nearby spiral galaxies by
the \emph{Einstein} satellite, a new class of intermediate luminosity ($L_X=10^{39}-10^{40}$
erg/s) X--ray sources was discovered (cf Fabbiano 1989). These sources, later defined as
ultraluminous X--ray sources (ULX, Makishima et al. 2000), were immediately intriguing, since
one of the proposed model is that they could be intermediate mass black holes ($10^{2}-10^{4}M_{\odot}$)
accreting at sub--Eddington rates, the missing link between stellar mass X--ray binaries
and active galactic nuclei (see Miller \& Colbert 2003 for a review on intermediate mass
black holes and their relationship with ULX). However, there are also other explanations available,
which do not require a new class of object. According to these models, the ULX are stellar mass
X--ray binaries, but either with truly super--Eddington accretion rate (e.g. Watarai et al. 2000, Begelman 2002),
or with sub--Eddington rate, but with some type of collimated emission, either simply anisotropic (King et al. 2001) or 
relativistic (K\"ording et al. 2002, Georganopoulos et al. 2002) to increase the observed luminosity. 
The threshold to define an ULX is now generally set to $10^{39.0}$ erg/s,
without any reference to the physical mechanism responsible for this value 
(see Miller \& Colbert 2003 for a review).

Surveys of ULX (e.g. with \emph{ROSAT} Colbert \& Ptak 2002, with \emph{XMM--Newton} Foschini et al. 2002,
with \emph{Chandra} Colbert et al. 2003) can
give gross information about these sources, their statistical properties, their relationships with the
host galaxy. However, to improve the understanding of these sources, a detailed study of 
nearby ULX with high signal--to--noise data are needed.

M33 (NGC 598) is one of the nearest spiral galaxies ($d=795$ kpc).
Classified as SA(s)cd, it has an inclination angle of $55^{\circ}$ (Ho et al. 1997). Since the
first observations with the {\it Einstein} satellite (Long et al. 1981), it was clear that the
central source (M33 X-8) had particular features (luminosity in the $0.2-4$ keV energy range
of about $10^{39}$ erg~s$^{-1}$, soft spectrum, excess of absorption along the line of sight)
suggesting that the source is somewhat different from an active galactic nucleus (Trinchieri et al. 1988).
The authors suggested the possibility that M33 contains a new type of X-ray binary system.

Later on, {\it ASCA} (Takano et al. 1994) observations extended up to 7 keV and
strengthened the early results of Trinchieri et al. (1988). The best-fit model
was composed of a multicolour disc (MCD) plus a power law at high energies, consistent
with that of Galactic black holes in their high state. However, Schulman \&
Bregman (1995), based on {\it ROSAT} observations, conclude that the probability
of such an unusual X-ray binary close to the centre of M33 is very small.

Another point which makes M33 X-8 an unusual source is the steadiness of its flux,
except for a modulation of $\sim 20\%$ with a period of 106 days (Dubus et al. 1997).
This discovery strengthened the hypothesis of a binary system, since the
modulations can
be due to the precession motion of the accretion disc (cf. Maloney et al. 1996).
It is important to add that there is a lack of information at wavelengths other than X-rays
for the source, since the source is located in a crowded region, so that it is
difficult to find the right counterpart or the companion star.

The recent increase of interest for ultraluminous X-ray source phenomenon gave new
light to the study of M33 X-8. Indeed, since the spatial resolution of
{\it Einstein},
{\it ROSAT}, and {\it ASCA} were not sufficient to rule out the possibility of a small offset of
the source from the optical centre, Makishima et al. (2000) suggested that X-8 could be an ULX.
However, {\it Chandra} observations put the tightest constraints on the position of X-8
(Dubus \& Rutledge 2002). The authors found a possible counterparts at radio wavelengths: it was 
identified with the point source n. 102 discovered by (Gordon et al. 1999) with the VLA at 20 and 6 cm. 
In the near-IR, M33 X--8 is at the 2MASS position of the nucleus (2MASS~J$01335089+3039365$)
within $0.6''$, which corresponds to about $2.3$ pc at the distance of $795$ kpc (Dubus \& Rutledge 2002).

The hypothesis of an active galactic nucleus (AGN) in the centre of M33 is inconsistent with the upper
limits on the central black hole mass obtained from the velocity dispersion measurements of the nuclear region:
Kormendy \& McClure (1993) gave an upper limit of $5\times 10^{4}\,M_{\odot}$, by
using the Canada-France-Hawaii Telescope. Recently, Gebhardt et al. (2001)
set, with the {\it Hubble Space Telescope (HST)}, an upper limit to only
$1500\,M_{\odot}$. Moreover the 106 days periodicity is not consistent
with the AGN hypothesis. The possibility that M33 X-8
is an ULX is the best explanation, as already suggested by Makishima et al.
(2000), although the source is very close to the centre of M33.

We present a detailed analysis of observations of the source M33 X-8 with
\emph{XMM-Newton}. This work is organized as follows: after the introduction
(Sect.~1), the X-ray data reduction and analysis are described in the Sect.~2.
Section~3 deals with the observations of the nuclear region of M33 in the
near-IR and radio wavelengths. The interpretation of the X-ray data is divided in the Sections~4 and 5:
the evaluation of the mass of the compact object is extensively dealt with in the first
part, while the second discusses the main characteristics of the source.

\begin{table*}[!ht]
\caption{\emph{XMM-Newton} Observation Log. Columns: (1) Observation Identifier;
(2) Date of the observation; (3) Duration of the observation [s]; (4,5,6) Observing mode
of MOS1, MOS2, and PN, respectively [FF: Full Frame; SW: Small Window]; (7) Position
with respect to the centre of the field of view.}
\centering
\begin{tabular}{lllllll}
\hline
ObsID & Date & Exposure & MOS1 & MOS2 & PN  & Position\\
(1)   & (2)  & (3)      & (4)  & (5)  & (6) & (7)\\
\hline
0102640101 & 04 Aug 2000 & $18672$ & SW & SW & FF & on axis\\
0102640301 & 07 Aug 2000 & $14862$ & FF & FF & FF & off axis\\
0102640601 & 05 Jul 2001 & $12525$ & FF & FF & FF & off axis\\
0102641001 & 08 Jul 2001 & $13275$ & FF & FF & FF & off axis\\
0102642001 & 15 Aug 2001 & $12266$ & FF & FF & FF & off axis\\
0102642101 & 25 Jan 2002 & $13001$ & FF & FF & FF & off axis\\
0102642301 & 27 Jan 2002 & $12999$ & FF & FF & FF & off axis\\
\hline
\end{tabular}
\label{tab:log}
\end{table*}

\section{\emph{XMM-Newton} observation and data reduction}
A set of observations of the central region of M33 is available in the 
\emph{XMM-Newton} Public Data Archive (see Table~\ref{tab:log}), with the 
nucleus in several position angles (on-axis and off-axis).
For the processing, screening, and analysis of the data from the EPIC MOS1 and
MOS2 cameras (Turner et al. 2001) and PN camera (Str\"uder et al. 2001),
we used the standard tools of XMM-SAS software v. 5.4.1 and
HEAsoft Xspec (11.2.0) Xronos (5.19) and followed the standard procedures
described in Snowden et al. (2002). In some cases, the observations were
affected by solar soft-proton flares, so that a preliminary 
cleaning was necessary.

\subsection{Time analysis}
To study the evolution of M33 X$-$8 and check for possible state
transitions, we extracted from the observations reported in Table~\ref{tab:log}
EPIC-PN light curves with $\sim 73$~ms time resolution. We extracted the data
from a circle with $35''$ radius and centered in the position of M33 X-8
($RA=01:33:50.89$, $Dec=+30:39:37.2$, J2000, uncertainty $<4''$).

\begin{figure}[!t]
\centering
\includegraphics[scale=0.35]{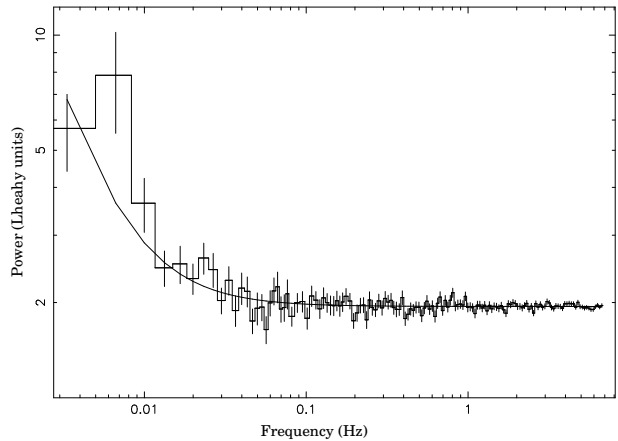}
\caption{Power density spectrum of M33 X-8 build from all the \emph{XMM-Newton} observations,
except for ObsID 0102640101, which showed instrumental noise. }
\label{pdspec}
\end{figure}

\begin{figure}[!t]
\centering
\includegraphics[scale=0.35]{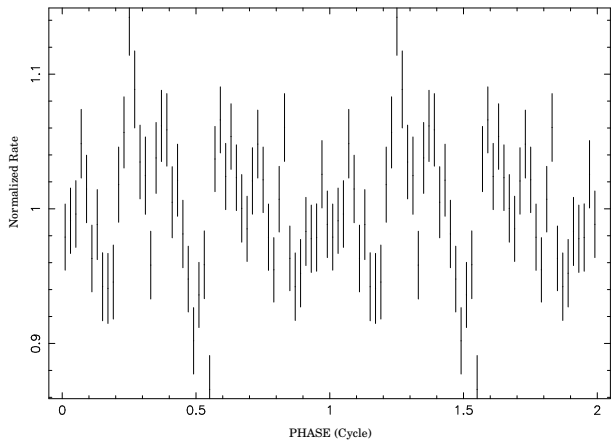}
\caption{Global light curve of M33 X-8 (all the observations) folded with a
period of $5000$~s.  Error bars are at $1\sigma$.}
\label{folding}
\end{figure}

The background was derived from a region $2'$ wide near the source in the PN camera.
Except for the third observation, during which the soft-proton flares
limit the good time to about 5 ks, the others had a net exposure time of
$\sim 9-10$ ks.

From all these light curves, but one, we produced power density spectra (PDS)
on interval of $\sim 300$~s, and all the resultant PDS of a single observation were averaged
and Leahy normalized (cf Leahy et al. 1983). The resultant PDS are thus fitted between
$3.3$~mHz and $6.8$~Hz (Fig.~\ref{pdspec}). Although the PDS above $\sim 20$~mHz is flat
and compatible with white noise, evidence for a red-noise component is found below
that value: our best fit with a power law gives an index $-1.5_{-0.2}^{+0.3}$
($\chi^2=130.6$, $137$~\emph{d.o.f.}). The red noise that appears to be
present is apparently not due to the source.

Given te present statistics, it is not possible to have sufficient
frequency resolution below $3.3$ mHz to study the signal at $0.2$ mHz ($5000$ s)
reported by La Parola et al. (2003), but the folding of the light curve on
this timescale indicates the presence of signal (see Fig.~\ref{folding}).
Although the exposure time is not sufficient to have a highly significative detection,
the $\chi^2$ test gives a probability of $4$~\% for the constancy of the source, 
with an excess variance $<2.1$~\% ($3\sigma$), thus confirming the results obtained by La Parola et al. (2003).

No state transitions were observed: we detect only a flux increase in the observation of
15 August 2001, without significant spectral changes with respect to the best-fit model
described in the Sect.~2.1, and the flux variation was consistent with the
well-known modulation of $\sim20$\% already observed by early satellites.

\subsection{Spectral analysis}
For the spectral analysis, we retrieved the on-axis observation of M33
performed on August $4^{\rm th}$, 2000.
The EPIC MOS cameras were set in the small-window mode, and the net exposure was
$12.5$~ks long. The PN camera was in full-frame mode, and the exposure was $13.5$~ks long.
We extracted the data from the same regions described in the Sect.~2.1.
The background was derived from a region $2'$ wide near the source in the PN camera, but for
the MOS cameras, since they operated in small-window mode, we used the background in the closest chip.
The spectra were rebinned so that each energy bin contained a minimum of 30 photons, and we
fitted only in the $0.5-10$~keV energy range because of the uncertainties in the MOS cameras calibration
at low energies (cf. Kirsch 2003). The photon redistribution matrix and the related ancillary
file were created appropriately with the \texttt{rmfgen} and \texttt{arfgen} tasks of XMM-SAS.

\begin{table*}[!t]
\caption{Results from the fit of the X-ray data. Columns: (1) Model: power law (PL),
power law with high-energy cutoff (COPL; \texttt{cutoffpl} model in \texttt{xspec}),
multicolour black body disc (MCD; \texttt{discbb} model in \texttt{xspec}), Raymond-Smith
(RAY), unsaturated Comptonization (CST; \texttt{compst} model in \texttt{xspec});
(2) Absorption column [$10^{21}$~cm$^{-2}$]; (3) free parameters of the model: (PL)
photon index $\Gamma$; (COPL) photon index $\Gamma$ and cutoff energy $E_{cut}$ [keV];
(MCD) temperature [keV] at the inner disc ($T_{in}$); (RAY) plasma temperature
[keV] and metal abundances $a$; (CST) temperature [keV] and optical depth $\tau$; (4)
$\chi^2$ and degrees of freedom of the spectral fitting; the reduced $\chi^2$
is reported between brackets; (5) flux in the $0.5-10$ keV band
($10^{-11}$~erg cm$^{-2}$ s$^{-1}$); (6) X-ray luminosity in the
$0.5-10$ keV band ($10^{39}$ erg~s$^{-1}$) calculated for $d=795$~kpc and corrected for the
absorption. The Galactic column is $N_{\mathrm H} =5.6\times 10^{20}$ cm$^{-2}$.
The uncertainties in the parameters are at the 90\% confidence level.}
\centering
\begin{tabular}{llllll}
\hline
Model & $N_{H}$ & Parameters & $\chi^2$/d.o.f. & $F_{X}$ & $L_{X}$\\
(1)   & (2)     & (3)        & (4)             & (5)     & (6) \\
\hline
MCD+PL   & $1.8\pm 0.2$      & $\Gamma=2.5_{0.1}^{0.2}$         & 1221.5/1175    & 1.7  & 1.7\\
{}       & {}                & $T_{in}=1.16\pm 0.04$~keV        & (1.04)         & {}   & {} \\
MCD+COPL & $1.4_{0.4}^{0.5}$ & $\Gamma=1.8_{0.8}^{0.7}$         & 1219.4/1174    & 1.7  & 1.5\\
{}       & {}                & $T_{in}=1.2_{0.1}^{0.2}$~keV     & (1.04)         & {}   & {}\\
{}       & {}                & $E_{cut}>2.3$~keV                & {}             & {}   & {}\\
RAY+COPL & $1.0_{0.8}^{0.2}$ & $\Gamma < 0.8$                   & 1216.8/1173    & 1.7  & 1.4\\
{}       & {}                & $a<0.97$                         & (1.04)         & {}   & {}\\
{}       & {}                & $kT=1.3_{0.3}^{0.5}$~keV         & {}             & {}   & {}\\
{}       & {}                & $E_{cut}=1.8_{0.2}^{0.3}$~keV    & {}             & {}   & {}\\
CST      & $1.65\pm 0.07$    & $kT=1.13\pm 0.03$~keV            & 1244.6/1176    & 1.7  & 1.6\\
{}       & {}                & $\tau=20.9\pm 0.6$               & (1.06)         & {}   & {}\\
\hline
\end{tabular}
\label{tab:xdata}
\end{table*}

\begin{figure}[!t]
\centering
\includegraphics[angle=270,scale=0.35]{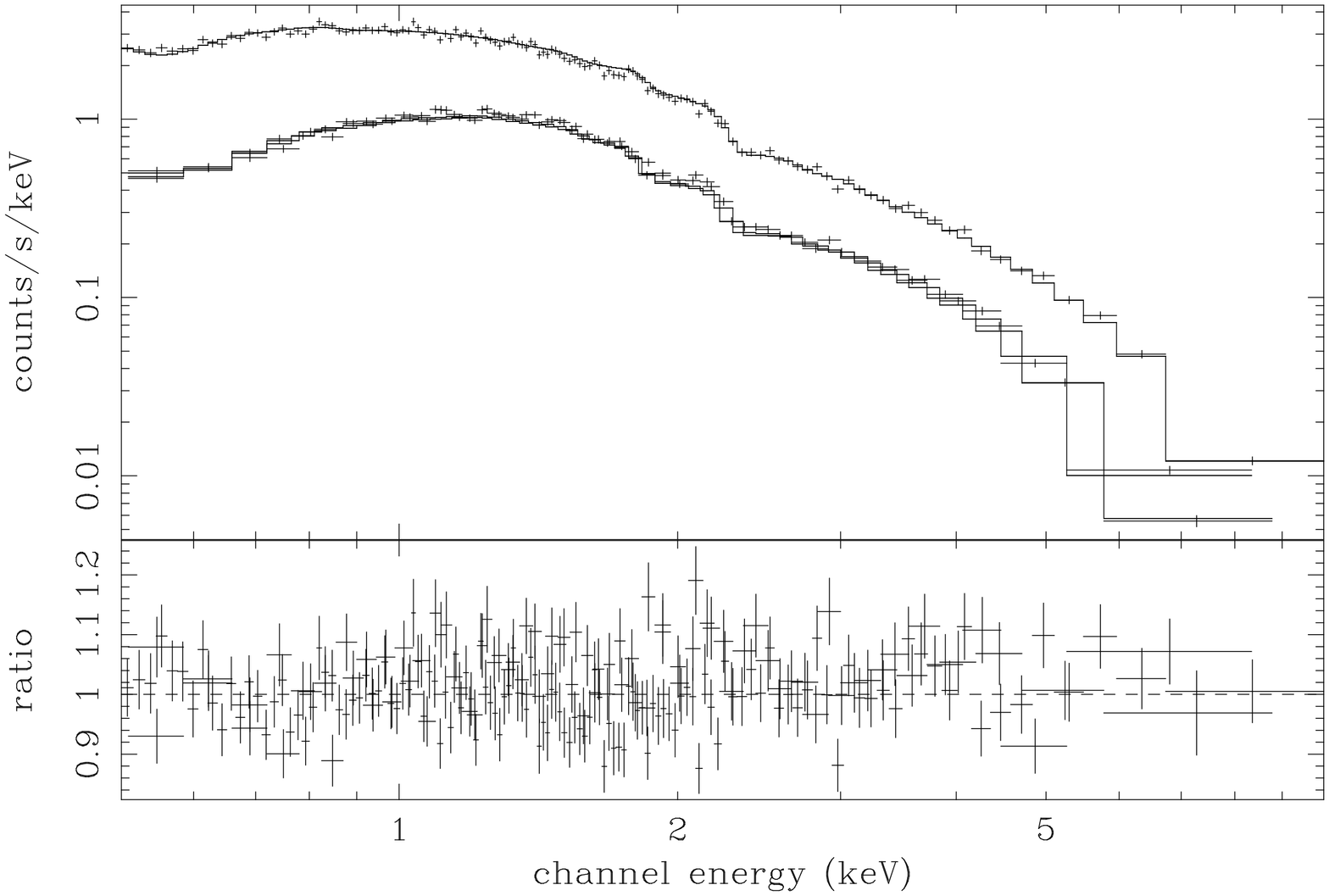}\\
\includegraphics[angle=270,scale=0.35]{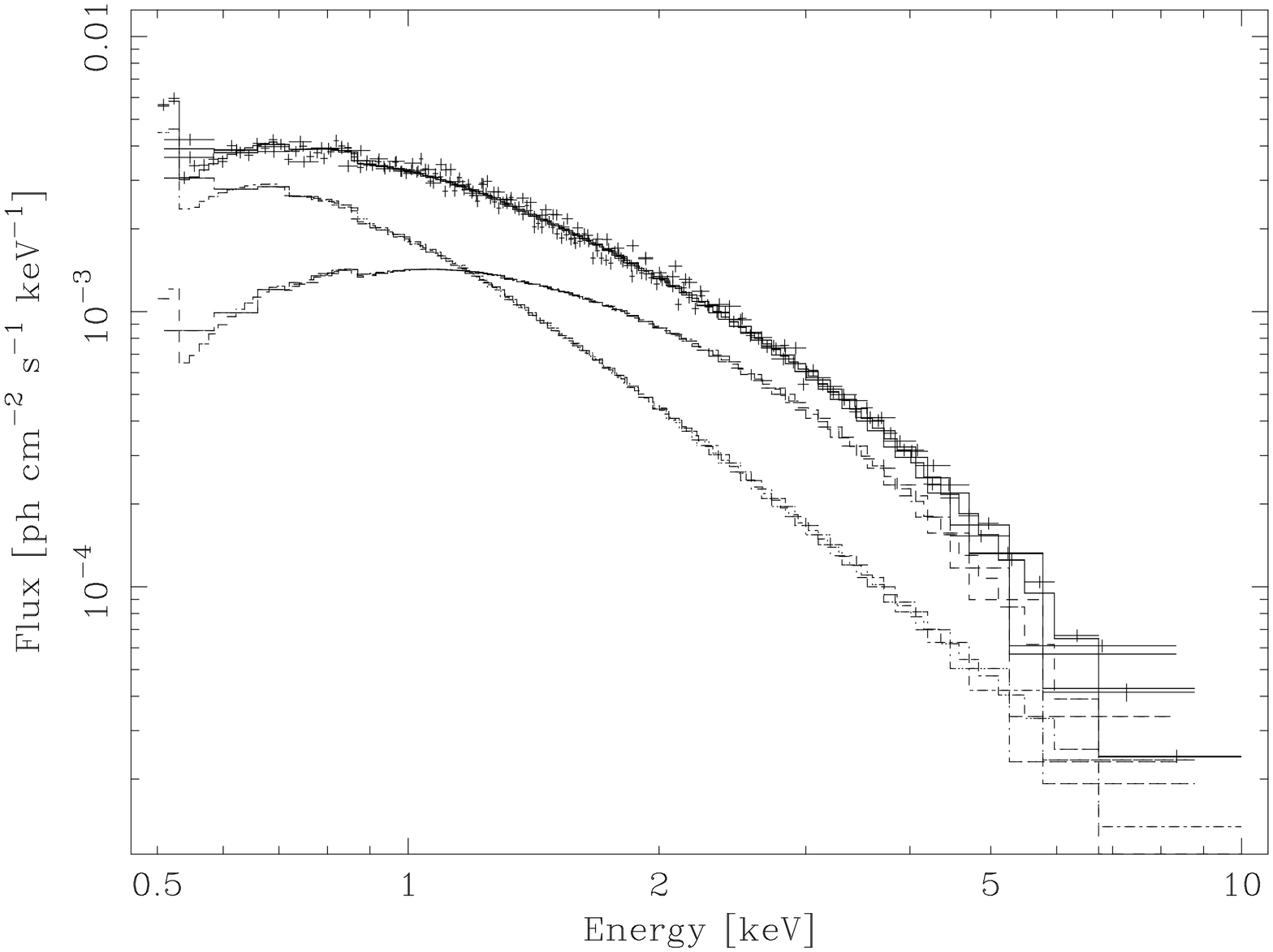}
\caption{$(top)$ Best-fit spectrum of M33--X8 with EPIC MOS1, MOS2, and PN data. The model
used is the multicolour disc plus a power law. See Table~\ref{tab:xdata} and the
text for details. The ratio data/model is shown in the bottom panel. $(bottom)$
Corresponding unfolded spectrum.}
\label{spec1}
\end{figure}

We tried to fit the spectrum obtained with EPIC PN and MOS cameras with several models.
Results of this are reported in Table~\ref{tab:xdata}. The best-fit model is composed
of a multicolour accretion disc with temperature at the inner disc $T_{in}=1.16$~keV plus
a power law with $\Gamma=2.5$ (Fig.~\ref{spec1}). The flux in the band $0.5-10$ keV is
of $1.7\times 10^{-11}$~erg cm$^{-2}$ s$^{-1}$ and is in agreement with earlier
observations of other satellites. The power law accounts for about $57$\%
of the total flux.

The absorption column is higher than the Galactic value of $N_{H}=5.6\times 10^{20}$~cm$^{-2}$
along the direction of observation, the latter being evaluated according to
Dickey \& Lockman (1990). In past observations, intrinsic absorption was never
mandatory: Schulman \& Bregman (1995) with {\it ROSAT} and Dubus \& Rutledge (2002) with {\it Chandra}
found that no absorption was required in addition to the Galactic hydrogen column.
On the other hand, Gottwald et al. (1987) with EXOSAT, Trinchieri et al. (1988)
with {\it Einstein}, Takano et al. (1994) with {\it ASCA}, Parmar et al. (2001) 
with {\it BeppoSAX}, and La Parola et al. (2003) with {\it Chandra} found that 
it was necessary to include an additional absorption component.

\begin{figure}[!t]
\centering
\includegraphics[angle=270,scale=0.35]{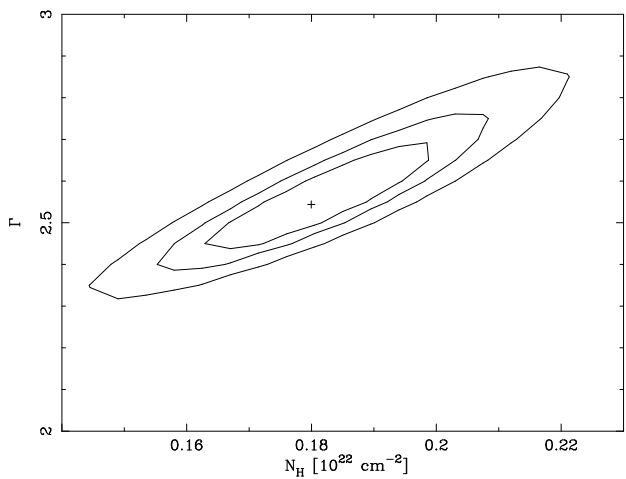}
\caption{2-dimensional fit-statistic contour plot of $\Gamma$ versus $N_{H}$.
The contours are with $\Delta \chi^2=2.3, 4.61, 9.21$, corresponding to the
confidence levels of $68$\%, $90$\%, and $99$\%.}
\label{spec2}
\end{figure}

In the present case, the additional absorption is required with
statistical significance greater than $99.99$\% (see Fig.~\ref{spec2} for the
2-dimensional fit-statistic contour plot of the power-law photon index and
the absorption column). The absorption along the line of sight appears
to be the same for both the multicolour disc and the power law model.

It is most probable that the earlier negative detections were due to
low statistics, rather than other effects: indeed, the {\it Chandra} spectrum with no
absorption (Dubus \& Rutledge 2002) was 10 ks long and had 23 degrees of freedom;
the observation from which La Parola et al. (2003) found additional absorption
was 92 ks long and had 333 degrees of freedom. It is worth noting too that the better
statistics obtained thanks to the large collecting area of \emph{XMM-Newton} results
in a smaller error range with respect to the previous measurement of the additional
absorption. The measured value of $1.24\times 10^{21}$~cm$^{-2}$ (already
subtracted for the Galactic value) corresponds to an optical
reddening of $E(B-V)=0.21$ mag, in agreement with the latest {\it HST}
observations that found $E(B-V)=0.22$ mag (Long et al. 2002).

La Parola et al. (2003) found that they needed to add a thermal plasma component.
However, substituting the multicolour disc model with a thermal plasma model
(e.g., Raymond-Smith) leads to a worse result, with parameters not properly constrained.
If the thermal plasma model is added to, instead of substituting, the MCD, the
results (not reported in Table~\ref{tab:xdata}) are even worse.

Furthermore, using a power law with an exponential cutoff, a model successfully used by some
authors (e.g., Gottwald et al. 1987; Trinchieri et al. 1988) did not improve the
fit, and in this case some parameters are also not properly constrained.

The only real alternative model to the reported best fit appears to be the unsaturated
Comptonization model of Sunyaev \& Titarchuk (1980). The plasma temperature is compatible
with that obtained from the multicolour accretion disc and the optical depth $\tau$, which is
known to vary according to the disc inclination ($\theta$), is compatible to a high value
$\theta\approx 60^{\circ}$. This agreement is expected in the case
of steady accretion discs around black hole candidates, as shown by Ebisawa et al. (1991).

There is, however, no evidence of any anomalous Comptonization as found by Kubota et al. (2001)
in GRO J$1655-40$. By adding to the best-fit model a Comptonized blackbody component (\texttt{compbb}
model in \texttt{xspec}) and linking the blackbody temperature to the temperature of the inner
disc ($T_{\mathrm{in}}$), the new three-component model does not converge.

We tried also the bulk motion Comptonization model (\texttt{bmc} model in \texttt{xspec},
Laurent \& Titarchuk 1999), which has been used successfully to fit the soft state of several
Galactic black hole candidates (Borozdin et al. 1999), but also some ULX, with M33 X--8 among
them (Schrader \& Titarchuk 2003); however, in the present case, the fit gives unphysical
results, with pegged parameters. We therefore do not mention this in Table~\ref{tab:xdata}.

Having detected flux variations in the observation of 15 August 2001, we extracted
the source spectrum to investigate the possibility of state transitions. The
data were affected by pile-up, mildly for the PN and strongly for the MOS cameras.
Therefore, we analyzed the data from PN only, and we extracted the source spectrum
from an annulus centered on the nucleus coordinate and with radii $10''$ and $40''$,
thus excluding the central region affected by pile-up. The background was extracted
from a nearby region of $2'$ radius.

The data were fitted to the best-fit model, i.e. multicolour accretion disc and power law.
The present fit gave $\chi^2=182$ for $d.o.f.=191$, with $N_{H}=(1.9\pm0.5)\times 10^{21}$~cm$^{-2}$,
$\Gamma=2.3_{-0.4}^{+0.8}$, and $T_{in}=1.4_{-0.4}^{+0.3}$~keV. Their results are consistent with the reference
spectrum within the measured errors. No state transition was observed, but the measured unabsorbed
flux in the energy band $0.5-10$~keV is $2.8\times 10^{-11}$~erg~cm$^{-2}$ s$^{-1}$, which is
$27\%$ higher than the flux measured during observation $0102640101$, consistent with
the well-known modulation of $\sim 20\%$.

\section{Observations of the nuclear region at other wavelengths}
Observations of the nuclear region of M33 at wavelengths other than X-rays
are very difficult because the stellar density in the nuclear region of M33 is
so high.  Even with the highest available resolution, the innermost
region of the nucleus remains unresolved (Lauer et al. 1998).

{\it Chandra} observations by Dubus \& Rutledge (2002) placed the X-ray
position within $0.6''$ ($2.3$~pc at the distance of $795$~kpc) of the near-IR
position from 2MASS and the radio position from the VLA observations. The {\it Chandra} coordinates
are compatible with what has been found by {\it ROSAT} and the present observation with \emph{XMM-Newton}.

The reference radio observation has been performed by Gordon et al. (1999). They observed
M33 with the VLA at 6 and 20 cm  with $7''$ resolution and detected the centre
of M33 at $RA=01:33:50.89$ and $Dec=+30:39:37.33$ (J2000). The flux density at 20 cm was $S_{20}=0.6\pm0.1$ mJy,
while it was $0.2\pm 0.1$ mJy at 6 cm (signal-to-noise ratio~$> 3$). The spectral index
was $\alpha=-0.8\pm 0.2$, where $S_{\nu}\propto \nu^{\alpha}$. By comparing
the radio data with optical observations performed with the $4$ m telescope at the Kitt Peak Observatory,
Gordon et al. (1999) exclude the possibility that the M33 centre is related to supernova remnants or H~II regions.
Intermediate-age stars, common in the nuclear region (Lauer et al. 1998),
should not generate strong radio emission;
the only other type of source that can display an $\alpha=-0.8$ spectrum would
be a background AGN. But the periodicity of 106 days in X-rays excludes the possibility of an AGN.
Therefore, it is reasonable to assume that the VLA detection is genuinely
linked with M33 X-8, although the current spatial resolution of either the
X-ray or radio data cannot yet definitively prove this.

However, there are problems to accept the 2MASS detection as simply the \emph{counterpart} of X-8.
In the 2MASS All Sky Data Release Catalog\footnote{http://www.ipac.caltech.edu/2mass/releases/allsky/.
The Two Micron All Sky Survey (2MASS) is a joint project of the University of
Massachusetts and the Infrared Processing and Analysis Center/California Institute of
Technology, funded by the National Aeronautics and Space Administration and the
National Science Foundation.} (released on March 2003) the centre of M33 is located at $RA=01:33:50.9$ and
$Dec=+30:39:36.6$ (J2000, spatial resolution $3''$). The apparent magnitudes are $J=12.06\pm 0.03$,
$H=11.44\pm 0.03$, and $K=11.22\pm 0.03$. The detection flags of the catalog indicate a good quality processing
of a pointlike source, although $3''$ at $795$ kpc is equivalent to about $11$ pc.

The total absorption column measured by the present work ($N_{H}=1.8\times 10^{21}$ cm$^{-2}$)
allows us to calculate a visual extinction of $A_{V}=N_{H}\cdot 5.3\times10^{-22}=0.954$ mag
(by using $R_{V}=3.1$; cf. Cox 2000). Then, it is possible to calculate the extinction factors at
the 2MASS wavelengths according to Cardelli et al. (1989): $A_{J}=0.27$, $A_{H}=0.18$, and $A_{K}=0.11$ mag.
The dereddened magnitudes are $J=11.79$, $H=11.26$, and $K=11.11$, which yield
colours of $J-H=0.53$ mag and $H-K=0.15$ mag, indicating an infrared excess that cannot be due to a single
star. The near-IR emission is consistent with the light expected from the
known nuclear star cluster of M33. According to Kormendy \& McClure (1993),
the central cluster has a $B$-band magnitude of $\sim 14.6$.  Long et al.
(2002) recently concluded that M33's nucleus has an age of
$\sim 10^7-10^9$ yrs.  From the population synthesis models of
Bruzual \& Charlot (2003), we anticipate that a cluster in this age range
should have $B-V \approx 0 - 0.6$ mag and $V - K \approx 2 - 3$ mag.
Hence, the nuclear star cluster of M33 is expected to have a $K$-band magnitude
of $\sim 11 - 12$, consistent with the value seen by 2MASS. Therefore, the
2MASS detection is likely to be dominated by the integrated stellar emission
of the nuclear star cluster, not by near-IR emission intrinsic to X-8.

\section{Discussion}
The source shows interesting similarities with other well known Galactic BHC
(see, e.g., McClintock \& Remillard 2003). The radio spectral index  of M33 X--8 is very 
similar to the ones of the famous microquasars 1E$1740.7-2942$ (cf Mirabel et al. 1992) 
or SS433 (cf Dubner et al. 1998). Moreover we confirm the $5000$~s modulation
of the emission of the source. If this period refers to a signal propagating at
the speed of sound (with a typical value of $c_s=10$ km~s$^{-1}$) or at the speed 
of light, the corresponding physical dimensions are between $5\times 10^{9}$ and 
$1.5\times 10^{14}$~cm. A similar variability has been observed also in SS433 
(jet $1000$~s) and for this source the typical dimension has been 
calculated to be about $10^{13}$~cm (Kotani et al. 2002).

Moreover the source X--ray flux recorded by \emph{XMM--Newton} is almost equal
to that previously measured. The only change observed is in good agreement
with the 106 days modulation previously measured (Dubus et al. 1997). By this evidence, 
we may assess that the source has been observed to be almost stable during the last
20 years. No particularly strong variation in the spectral and/or flux state
of the source has ever been observed. These characteristics are very 
similar to what is observed in the BHC LMC X--1 (cf Nowak et al. 2001, Wilms et al. 2001).

From these indications we are led to classify the ULX M33 X--8 as a black hole
with stellar mass. This picture is supported also by the X--ray spectra we 
obtained with \emph{XMM--Newton}. In particular, the temperature of the thermal 
component seems to be too high to be referred to an intermediate mass BH unless one assumes
relativistic effects (see the next subsection).

\subsection{Evaluation of the mass of M33 X-8}
To calculate the mass of M33 X-8, we have to take into account that it cannot
be greater than the upper limit of the non-luminous mass of the nucleus,
$1500\,M_{\odot}$ (Gebhardt et al. 2001).

From the X-ray analysis, it is obvious that X-8 is in a very high state, with 
a thermal component with $kT\sim 1$ keV and a power law with photon index $\sim 2.5$. 
According to some authors (e.g. Done \& Gierlinski 2003), these
spectral characteristics are the typical signature of the accretion disc around a black hole.

In the present work, the best model to fit the ultrasoft component is the multicolour
accretion disc (MCD) by Mitsuda et al. (1984). The MCD model require, to be correctly used, 
some additional parameters (cf Merloni et al. 2000, Ebisawa et al. 2003). Therefore, we recall 
some basic definitions to explain the values of the parameters we used to calculate the mass 
of the compact object in the present case. We refer to the works of Makishima et al. (2000) 
and Ebisawa et al (2003) for further details and deeper analysis on the MCD model applied to ULX.

The normalization of the MCD model $A_{\mathrm{MCD}}$ allows a direct estimate
of the inner disc radius $R_{\mathrm{in}}$, by means of $A_{\mathrm{MCD}}=R_{\mathrm{in}}^2 \cos \theta/D^2$, where
$D$ is the distance of the source in units of $10$ kpc, $\theta$ is the inclination of the
disc ($\theta=0^{\circ}$ means face-on; $\theta=90^{\circ}$ refers to
the disc edge-on). $R_{\mathrm{in}}$ is expressed in km and depends on the spin of the black hole. In the case of
a Schwarzschild black hole (spin 0), $R_{\mathrm{in}}$ is equal to three times the event horizon radius
$R_{\mathrm{S}}=2GM/c^2$ (that is twice the gravitational radius), while for a Kerr black hole, the radius of the inner
disc can be down to $1.24R_{\mathrm{S}}$ in the most extreme case of spin $+1$. 

Two more correction parameters should be taken into account: the first
correction, indicated with the parameter $\xi$, is to represent the fact that
$T_{\mathrm{in}}$, the temperature at the innermost disk boundary, is related to a radius a 
bit larger than $R_{\mathrm{in}}$ ($\xi\approx 0.41$,
Kubota et al. 1998). The second parameter is the spectral hardening factor $f$ of
Shimura \& Takahara (1995), which takes into account the fact that in the 
MCD model $T_{\mathrm{in}}$ is the maximum disc colour temperature, and therefore it has to 
be converted into the effective temperature. The hardening factor weakly depends on the 
accretion rate and the viscous parameter $\alpha$ of the standard model (Shakura \& Sunyaev 1973). 
For Galactic black holes, $f$ is generally constant and within the range $1.7-1.9$ 
(see Ebisawa et al. 2003 for a discussion on these values). We assume $f=1.7$.

The mass of the compact object is therefore given by $M=(R_{\mathrm{in}}\xi f^2/8.86 s)\,M_{\odot}$, 
where $s$ is a coefficient depending on the spin of the black hole. In the case of a
Schwarzschild black hole $s=1$. A further uncertainty in the evaluation of the mass is given by the
inclination of the accretion disc, which is generally unknown. By assuming $\cos \theta = 1$,
we obtain a lower limit of the mass. The MCD normalization in the best-fit model of M33 X-8 
gives $R_{\mathrm{in}}=46\pm 3$~km, which corresponds to a mass of $(6.2\pm 0.4) \,M_{\odot}$. 

It is worth noting that the inclination of the accretion disc and the
corresponding relativistic corrections can increase the value of the mass.
To have an estimate of the possible inclination of the accretion disc,
we note in Table \ref{tab:xdata} that the X--ray spectrum of M33 X--8 is also well fitted
by the unsaturated Comptonization model by Sunyaev \& Titarchuk (1980), with a temperature
compatible with that of the MCD model. Therefore, it is reasonable to think that
the power law component of the MCD+PL model could be due to the Comptonization.
From the data of fit with unsaturated Comptonization and the studies performed by
Ebisawa et al. (1991), we can infer that a possible value for
$\theta=60^{\circ}$. If true, the mass would be $M=(8.7\pm 0.5)\,M_{\odot}$.

If we include also relativistic effects of the disc inclination, including
Doppler boosting and gravitational focusing, by assuming that the accretion
disc shares the same behavior as that seen in AGNs (see Sun \& Malkan 1989),
the mass might rise to $M=(12\pm 1)\,M_{\odot}$. 

Defining $L_{\rm Edd}=1.3(M/M_{\odot})\times 10^{38}$ erg/s 
(e.g. Frank et al. 2002), and assuming $L_{0.01-100 \rm keV}=2.7\times 10^{39}$~erg/s
(extrapolated from the X--ray data) as the best approximation of the bolometric luminosity, 
it results that we are observing the source at about $1.7\,L_{\rm Edd}$ for a $12\,M_{\odot}$ 
black hole. To reach such a super--Eddington luminosity there are two possibilities: to
have genuine super--Eddington accretion rate or sub--Eddington rate with some type of
anisotropies or collimation. The bolometric luminosity of the accretion disc only can
be calculated by using the Stefan--Boltzmann law with the dimension and temperature of
of the innermost stable region. In the case of the inclination angle $\theta=60^{\circ}$,
the disc can account for $L_{\rm disc}=1.0\times 10^{39}$ erg/s. The remaining part of the 
accretion, related to the power law component, is more difficult to explain, since it
depends on the physical interpretation. In the case of sub--Eddington rate,
a little anisotropy (solid angle of the emission $<1.6\pi$) is sufficient to account for the 
observed flux. These values are similar to those of the Galactic microquasar GRS$1915+105$, 
where the inclination angle is $70^{\circ}$ and the solid angle of the emission is about 
$2.4\pi$ (see King et al. 2001). We cannot exclude at all that the M33 X--8 could 
accrete at super--Eddington rates, although, in this case, we would expect values of
the excess variance greater than those observed. 

We caution that the mass evaluation is only as good as the MCD model, 
the present best fit, and the parameters adopted/inferred for the correction of the MCD.

The above value is the best estimate, according to the \emph{XMM--Newton}
observation and in the hypothesis of a Schwarzschild black hole. It is worth
noting in case of intermediate mass black hole, the temperature of the inner disc
is expected to be much lower than the $1$ keV measured in the present case. 
Miller et al. (2003) found a temperature of $kT_{\rm in}=0.15$ keV for the two 
ULX in NGC1313 from which they inferred the presence of a black hole of 
$\approx 10^3\, M_{\odot}$. It is possible to reach a high value of the mass only
if the black hole is maximally rotating ($s=1/6$, Kerr black hole).
In this case, taking into account all the effects and parameters described 
above, the mass of M33 X--8 could be as high as $M=(147\pm 8)\,M_{\odot}$.

\subsection{A mildly relativistic jet or a Compton heated wind?}
We propose two possible hypotheses to explain the power law component of M33 X-8.
The first is the presence of \emph{collimated emission (mildly relativistic jets)}.
The dimensions of the variable region appear to be compatible
with the X-ray emitting region from jets in microquasars: e.g. for SS433 (jet speed $0.26c$,
time variability $1000$ s), this region has dimension $\sim 10^{13}$ cm (Kotani et al. 2002).
The radio spectral index is compatible with the values found in the hot spots of some Galactic
microquasars, like 1E$1740.7-2942$ (Mirabel et al. 1992) or SS433 (Dubner et al. 1998).
Although it is not possible to resolve the radio emission from the nucleus of M33, the steep
spectral index is consistent with the synchrotron emission from charged particles accelerated
in shocks generated by the propagation of a jet in a diffuse region. 
But a cautionary note should be stressed, which is the great uncertainty in the determination 
of the counterpart of M33 X-8.

The second hypothesis takes into account the presence of \emph{Compton-heated winds},
like, for example, the Galactic black hole GRO~J$0422+32$ (van Paradijs et al. 1994). 
This model was never proposed to explain the high luminosity of ULX. In this case, 
the hard X-ray emission represents the signature of
a specific physical process, generally taken to be inverse-Compton scattering of photons
of thermal origin on a population of hot electrons. The geometry of the region where
this process occurs is rather difficult to understand. The two-phase model
by Haardt \& Maraschi (1991) of a corona in hydrostatic equilibrium around the
accretion disc is one of the standard solutions. In this case, thermal radiation emitted 
by the accretion disc enters the hot corona and is Comptonized into hard X-rays. Part of 
this radiation is then reprocessed by the accretion disc, and a small fraction is reflected. 
The fact that the spectrum of M33 X-8 is well fitted also by the unsaturated Comptonization 
model strengthens the importance of the corona for this source (cf. Table~\ref{tab:xdata}).

We consider, as a reference, the model developed by Begelman et al. (1983).
A nearly hydrostatic corona exists at a distance $r$ from the centre of the accretion disc 
system if the Compton temperature $T_{\mathrm{IC}}$ is less than the escape temperature. This 
occurs inside the radius  $R_{\mathrm{IC}}= (1.0\times 10^{10}/T_{\mathrm{IC8}})M/M_{\odot}$, 
where $R_{\mathrm{IC}}$ is in cm and $T_{\mathrm{IC8}}$ is the inverse-Compton
temperature expressed in units of $10^{8}$ K (Begelman et al. 1983). If $r>R_{\mathrm{IC}}$, Compton heating can
cause a strong wind (see Begelman et al. 1983, Begelman \& McKee 1983, Shields et al. 1986).
In this case, a wind-driven relaxation cycle is set up, causing oscillations in the interplay
of the disc accretion rate and the wind ejection rate. In the standard disc model, these
oscillations have a period $P= (3400 \ \mathrm{s})\cdot M^{14/9}/(\alpha^{7/9} T^{4/3}_{\mathrm{IC8}} \dot{M}^{1/3}_{\mathrm{a17}}$),
where $\dot{M}_{\mathrm{a17}}$ is the disc accretion rate in units of $10^{17}$~g~s$^{-1}$.
In the hypothesis that M33 X--8 is accreting at about $60\%$ the Eddington limit,
from the above equation, it is possible to calculate $T_{\mathrm{IC8}}$, which in turn gives us
$R_{\mathrm{IC}}$. For the values of mass of the compact object
$M=(6-12) \,M_{\odot}$, we infer a value of $R_{\mathrm{IC}}\approx 7\times 10^9$~cm. Therefore,
if the oscillation occurs at this distance, the perturbation speed is about $14$~km~s$^{-1}$, compatible
with the sound speed in a isothermal plasma at a temperature of about $2\times 10^4$~K.
This value can be compared with what has been found in the case of GRO~J$0422+32$,
where the temperature of the plasma is $\sim 3\times 10^4$~K (van Paradijs et al. 1994).

It is worth mentioning that outflows have been invoked to account for the high 
luminosity in ULXs (e.g., Begelman 2002; King 2002; King \& Pounds 2003) and
the Compton-heated winds solution proposed here can be considered a variant of these
models.

\section{Final remarks}
We presented the spectral and temporal analysis of \emph{XMM-Newton}
observations of M33 X-8. The present analysis of X-ray data suggest that M33 X-8
is a stellar mass black hole, whose luminosity is only apparently 
super--Eddington for geometrical reasons. The lower limit for the mass of $M>6\, M_{\odot}$,
and a best estimate of $M=12\pm 1\,M_{\odot}$, although we cannot completely exclude a
mass of $\approx 150\, M_{\odot}$ if X--8 is a maximally rotating BH. 
These conclusions are in agreement with the X--ray binary interpretation already found by several other 
authors (Makishima et al. 2000, Dubus \& Rutledge 2002, King 2002, La Parola et al. 2003, 
just to mention the latest).

We confirm the oscillation with a period of $5000$~s discovered by La Parola 
et al. (2003), and we suggest that this oscillation is associated with the 
interplay between the mass loss from a Compton-heated wind and the accretion rate.
It is worth noting that also a mildly relativistic jet could explain as well
most of the observed data. In this case, the $5000$ s variability is due
to oscillations at the basis of the jet.

The case of M33 X-8 sheds new light on ULX studies.  The interpretation
proposed here for M33 X-8 --- a stellar-mass black hole whose luminosity 
is boosted by orientation effects of the accretion disc and Compton-heated 
winds (or even a mildly relativistic jet) ---  may serve as a useful 
template for understanding other ULXs.  To date, ULXs with little or no variability 
have generally been associated with young supernovae remnants. M33 X-8 illustrates 
that steady X-ray sources, with weak short term variability, can be stellar-mass 
black holes. 

It is also of interest to note that environments rich in hot plasma,
coming from hot winds of young stars or from stellar collisions, as might
occur in compact young star clusters or the nuclei of galaxies, may be
particularly conducive to fueling and sustaining ULX sources.

Finally, we would like to emphasize that very high resolution, multiwavelength,
simultaneous observations of M33 X-8 are required to draw definitive conclusions 
on the nature of this enigmatic source.

\begin{acknowledgements}
LF and MD acknowledge partial financial support by the Italian Space Agency
(ASI).  JR and YF acknowledge financial support from the CNES.  LCH is
is supported by the Carnegie Institution of Washington and by NASA grants from
the Space Telescope Science Institute (operated by AURA, Inc., under NASA
contract NAS5-26555). LF wishes to thank Giorgio Palumbo, Paola Grandi,
and Massimo Cappi for useful discussions.  This publication has 
made use of public data obtained from the High Energy Astrophysics Science 
Archive Research Centre (HEASARC), provided by NASA Goddard Space Flight Centre.
\end{acknowledgements}

\end{document}